\documentclass{article}
\usepackage{amsmath, amssymb, amsfonts}
\title{On a free of centrifugal acceleration spacetime} 
\author{Hristu Culetu\\ Ovidius University, Dept.of Physics and Electronics, \\B-dul Mamaia 124, 900527 Constanta, Romania \\ email : hculetu@yahoo.com}
\begin{document}
\numberwithin{equation}{section}
\pagenumbering{arabic}
\maketitle
\begin{abstract}
A static spacetime with no centrifugal repulsion, previously studied by Dadhich, is investigate in this paper. The source of curvature is considered to be an anisotropic fluid with $\rho = -p_{r}$ and constant angular pressures. The positive parameter from the line-element is interpreted as the invariant acceleration of a static observer. The timelike and null geodesics of the spacetime are examined. A regularized form of the metric is proposed, rendering it finite at the origin. The energy density of the fluid becomes finite and negative for any $r$ and all the pressures are positive throughout the spacetime. The Tolman-Komar energy  $W(r)$ is computed and proves to be smaller than that one calculated with the Dadhich metric.  
\end{abstract}

\newcommand{\fv}{\boldsymbol{f}}
\newcommand{\tv}{\boldsymbol{t}}
\newcommand{\gv}{\boldsymbol{g}}
\newcommand{\OV}{\boldsymbol{O}}
\newcommand{\wv}{\boldsymbol{w}}
\newcommand{\WV}{\boldsymbol{W}}
\newcommand{\NV}{\boldsymbol{N}}
\newcommand{\hv}{\boldsymbol{h}}
\newcommand{\yv}{\boldsymbol{y}}
\newcommand{\RE}{\textrm{Re}}
\newcommand{\IM}{\textrm{Im}}
\newcommand{\rot}{\textrm{rot}}
\newcommand{\dv}{\boldsymbol{d}}
\newcommand{\grad}{\textrm{grad}}
\newcommand{\Tr}{\textrm{Tr}}
\newcommand{\ua}{\uparrow}
\newcommand{\da}{\downarrow}
\newcommand{\ct}{\textrm{const}}
\newcommand{\xv}{\boldsymbol{x}}
\newcommand{\mv}{\boldsymbol{m}}
\newcommand{\rv}{\boldsymbol{r}}
\newcommand{\kv}{\boldsymbol{k}}
\newcommand{\VE}{\boldsymbol{V}}
\newcommand{\sv}{\boldsymbol{s}}
\newcommand{\RV}{\boldsymbol{R}}
\newcommand{\pv}{\boldsymbol{p}}
\newcommand{\PV}{\boldsymbol{P}}
\newcommand{\EV}{\boldsymbol{E}}
\newcommand{\DV}{\boldsymbol{D}}
\newcommand{\BV}{\boldsymbol{B}}
\newcommand{\HV}{\boldsymbol{H}}
\newcommand{\MV}{\boldsymbol{M}}
\newcommand{\be}{\begin{equation}}
\newcommand{\ee}{\end{equation}}
\newcommand{\ba}{\begin{eqnarray}}
\newcommand{\ea}{\end{eqnarray}}
\newcommand{\bq}{\begin{eqnarray*}}
\newcommand{\eq}{\end{eqnarray*}}
\newcommand{\pa}{\partial}
\newcommand{\f}{\frac}
\newcommand{\FV}{\boldsymbol{F}}
\newcommand{\ve}{\boldsymbol{v}}
\newcommand{\AV}{\boldsymbol{A}}
\newcommand{\jv}{\boldsymbol{j}}
\newcommand{\LV}{\boldsymbol{L}}
\newcommand{\SV}{\boldsymbol{S}}
\newcommand{\av}{\boldsymbol{a}}
\newcommand{\qv}{\boldsymbol{q}}
\newcommand{\QV}{\boldsymbol{Q}}
\newcommand{\ev}{\boldsymbol{e}}
\newcommand{\uv}{\boldsymbol{u}}
\newcommand{\KV}{\boldsymbol{K}}
\newcommand{\ro}{\boldsymbol{\rho}}
\newcommand{\si}{\boldsymbol{\sigma}}
\newcommand{\thv}{\boldsymbol{\theta}}
\newcommand{\bv}{\boldsymbol{b}}
\newcommand{\JV}{\boldsymbol{J}}
\newcommand{\nv}{\boldsymbol{n}}
\newcommand{\lv}{\boldsymbol{l}}
\newcommand{\om}{\boldsymbol{\omega}}
\newcommand{\Om}{\boldsymbol{\Omega}}
\newcommand{\Piv}{\boldsymbol{\Pi}}
\newcommand{\UV}{\boldsymbol{U}}
\newcommand{\iv}{\boldsymbol{i}}
\newcommand{\nuv}{\boldsymbol{\nu}}
\newcommand{\muv}{\boldsymbol{\mu}}
\newcommand{\lm}{\boldsymbol{\lambda}}
\newcommand{\Lm}{\boldsymbol{\Lambda}}
\newcommand{\opsi}{\overline{\psi}}
\renewcommand{\tan}{\textrm{tg}}
\renewcommand{\cot}{\textrm{ctg}}
\renewcommand{\sinh}{\textrm{sh}}
\renewcommand{\cosh}{\textrm{ch}}
\renewcommand{\tanh}{\textrm{th}}
\renewcommand{\coth}{\textrm{cth}}

\section{Introduction}
The Einsteinean gravity is self-contained and links universally to all particles including photons. In contrast, Newtonian gravity affects massive particles but the massless ones behave as they felt no gravitation \cite{ND1}. Moreover, in the classical gravity it is impossible to cancel everywhere the centrifugal acceleration. As Dadhich \cite{ND2} has shown, the additional attractive force due to the coupling gravitational - centrifugal potential (a purely Einsteinean effect) counters the centrifugal acceleration everywhere. He also proved that particles can have angular velocity but no centrifugal radial repulsion. Dadhich further looks for the physical source which gives rise to his non-asymptotically flat space, thus introducing a global monopole of unit charge in an AdS spacetime. In addition, the free of centrifugal forces geometry is singular at the origin (the curvatures diverge there) but, however, $r = 0$ is a horizon of the spacetime.

Our purpose in this paper is to investigate the free of centrifugal acceleration geometry and to give a different interpretation of the nature of the source of the field. In our view, Dadhich's constant $k$ represents the invariant acceleration (squared) of a static observer. The energy-momentum tensor corresponds to an anisotropic fluid and its energy density and radial pressures change their sign at some distance from the origin of coordinates. We study further the timelike and null geodesics and find that the test particle reaches the singularity after an infinite time, following a spiral trajectory. We get rid of the singularity at $r = 0$ with the help of an exponential term in the metric coefficients which renders the curvature invariants and the anisotropic stress tensor regular everywhere.

\section{Anisotropic fluid stress tensor}
We write down the "`free of centrifugal acceleration"' spacetime as
  \begin{equation}
ds^{2} = -g^{2}r^{2} dt^{2} + \frac{dr^{2}}{g^{2}r^{2}} + r^{2} d\Omega^{2}
\label{2.1}
\end{equation}
where $d\Omega^{2} = d\theta^{2} + sin^{2} \theta d\phi^{2}$ stands for the metric on the unit 2 - sphere and $g$ is a positive constant. Our aim is to investigate the structure of the energy-momentum tensor that leads to the metric (2.1), namely we look for $T_{ab}$ which solves Einstein's equations $G_{ab} = \kappa T_{ab}$, where $G_{ab}$ is the Einstein tensor and $\kappa = 8\pi G/c^{4}$ is Einstein's constant. We take from now on $c = 8\pi G = 1$. One obtains
  \begin{equation}
  T^{t}_{~t} = -\rho = 3g^{2} - \frac{1}{r^{2}},~~~ T^{r}_{~r} = p_{r} = -\rho,~~~T^{\theta}_{~\theta} = T^{\phi}_{~\phi} = 3g^{2} \equiv p_{t}
\label{2.2}
\end{equation}
where $p_{r}$ is the radial pressure and $\rho$ is the energy density.
Eqs. (2.2) are rooted from the general expression of the energy - momentum tensor for an anisotropic fluid \cite{HC4, KM}
 \begin{equation}
T^{a}_{b} = (\rho + p_{t})u^{a} u_{b} + p_{t} \delta^{a}_{b}+ (p_{r} - p_{t}) s^{a} s_{b}.
\label{2.3}
\end{equation}
 In (2.3), $u^{a}$ is the timelike velocity vector of the fluid and $s^{a}$ is spacelike , on the direction of anisotropy, with $s^{a} u_{a} = 0$ and $s^{a} s_{a} = 1$.

 It is worth to notice that Dadhich \cite{ND2} separated a $\Lambda$-term from $T^{a}_{~b}$ and, therefore, in his model the transverse pressures are vanishing. We see from (2.2) that $p_{r} = -\rho$ everywhere but the angular pressures $p_{t}$ are constant and positive. In addition, $\rho$ and $p_{r}$ change their sign at $r = 1/\sqrt{3}g$ and $T^{a}_{~b}$ acquires a $\Lambda$-form when $r \rightarrow \infty$. In contrast, $\rho$ and $p_{r}$ are divergent at the central singularity $r = 0$ which represents a horizon of the spacetime. We mention that there is no any central mass and, therefore, the origin of coordinates is arbitrary. That was probably the reason why Dadhich used a unit charge global monopole at $r = 0$ to justify the form of the spacetime (2.1). The central singularity is rooted from the divergence of the scalar curvature and the Kretschmann scalar 
   \begin{equation}
   R^{a}_{~a} = -12g^{2} + \frac{2}{r^{2}},~~~~K = \frac{4}{r^{4}} (1 - 2g^{2}r^{2} + 6g^{4}r^{4})
\label{2.4}
\end{equation}
at $r = 0$. In addition, $K$ is regular at infinity, i.e. $K_{\infty} = 24g^{4}$.

Let us consider a congruence of static observers with the velocity vector field $u^{a} = (1/gr, 0, 0, 0)$. The acceleration 4 - vector is given by
  \begin{equation}
   a^{b} \equiv  u^{a} \nabla_{a} u^{b} = (0, g^{2}r, 0, 0)
\label{2.5}
\end{equation}
The radial component $a^{r} = g^{2}r$ from (2.5) is the acceleration our observer would exert on a static particle to maintain it at $r = const.$ Because $a^{r} > 0$, the gravitational field is attractive. From (2.5) we immediately obtain $\sqrt{g_{bc}a^{b} a^{c}} = g$. In other words, the constant $g$ from the metric (2.1) is nothing but the invariant acceleration of the static observer. This interpretation is supported by the fact that for $r >> 1/\sqrt{3}g$, the energy density $\rho \propto g^{2}$, as in Newtonian gravity, $g$ being here the intensity of the gravitational field. We also observe that the projection of $a^{b}$ on the radial direction is constant, too: $a^{b}n_{b} = g$, where $n^{b} = (0, gr, 0, 0)$.
 
 Let us compute now the Tolman-Komar (TK) gravitational energy. We have \cite{TP, HC}
   \begin{equation}
W = 2 \int(T_{ab} - \frac{1}{2} g_{ab}T)u^{a} u^{b} N\sqrt{\gamma} d^{3}x ,
\label{2.6}
\end{equation}
with $N = \sqrt{-g_{00}} = gr$ and $\gamma$ is the determinant of the spatial three-metric. By means of $T^{a}_{~b}$ from (2.2) and with the above $u^{a}$ we get
\begin{equation}
W = g^{2} r^{3}
\label{2.7}
\end{equation}
$W$ is finite throughout the spacetime (2.1) and vanishes at the singularity. Even though $\rho < 0$ in the region $r > 1/\sqrt{3}g$, $W$ is positive everywhere because the pressures carry positive contribution in (2.6). The same result (2.7) should have been obtained if we calculated the gravitational energy by means of the formula \cite{TP1}
   \begin{equation}
W = \frac{1}{4\pi} \int_{\partial V} N a_{b}n^{b} \sqrt{\sigma} d^{2}x ,
\label{2.8}
\end{equation}
where $\sigma$ is the determinant of the metric on the 2-dimensional boundary $\partial V$ of constant $r$ and $t$ and $n^{b} = (0, 1/\sqrt{g_{rr}},0 ,0)$ is the spacelike normal on $r$= const. hypersurfaces. 

\section{Geodesics}
a) \textbf{Timelike geodesics}\\
The spacetime (2.1) is static and spherically symmetric and, therefore, we have two Killing vectors, $\xi^{a}_{(t)} = (1, 0, 0, 0)$ and $\xi^{a}_{(\phi)} = (0, 0, 0, 1)$. Taking for convenience $\theta = \pi/2$, (2.1) yields
  \begin{equation}
 g^{2}r^{2} \dot{t}^{2} - \frac{1}{g^{2}r^{2}}\dot{r}^{2} - r^{2} \dot{\phi}^{2} = 1,
\label{3.1}
\end{equation}
where $\dot{t} = dt/d\tau$, etc. and $\tau$ represents the proper time. From $-E = g_{ab}u^{a}\xi^{b}_{(t)}$ and $L = g_{ab}u^{a}\xi^{b}_{(\phi)}$, one obtains
  \begin{equation}
  \dot{t} = \frac{E}{g^{2}r^{2}},~~~~\dot{\phi} = \frac{L}{r^{2}}
\label{3.2}
\end{equation}
where $E$ and $L$ are the energy per unit mass of the test particle and, respectively, the angular momentum per unit mass. The velocity field vector is given by  $u^{a} = (\dot{t}, \dot{r}, 0, \dot{\phi}$). From (3.1) we get the radial equation
  \begin{equation}
  \dot{r}^{2} = E^{2} - g^{2}L^{2} - g^{2}r^{2}
\label{3.3}
\end{equation}
We see that the constants of motion $E$ and $L$ appear on equal footing in (3.3), excepting the minus sign in front of $L$. Therefore, we denote $b = \sqrt{E^{2} - g^{2}L^{2}}$, ($E > gL$), a constant parameter. Moreover, we must have $r \leq b/g$, i.e. the particle motion is restricted to a finite range of $r$. Keeping in mind that $\dot{r} = (E/g^{2}r^{2})dr/dt$, one easily finds from (3.3) that 
  \begin{equation}
 -\frac{\sqrt{b^{2} - g^{2}r^{2}}}{b^{2}r} = \pm \frac{g^{2}}{E}t + C 
\label{3.4}
\end{equation}
with $C$ a constant of integration. Taking $r = r_{max} = b/g$ at $t = 0$ as initial condition, we have $C = 0$. The radial component of the equation of motion reads
  \begin{equation}
  r(t) = \frac{bE}{g\sqrt{E^{2} + b^{4}g^{2}t^{2}}},
\label{3.5}
\end{equation}
where the sign in front of $t$ from (3.4) was chosen so that $ dr/dt < 0$ (ingoing geodesics) because of the initial condition used. The radial geodesics is obtained from (3.5) if we replaced $b$ with $E$ ($L = 0$). The last equation shows that the test particle needs an infinite time to reach the central singularity. The curve $r(t)$ has inflexions at $t_{i} =\pm E/\sqrt{2}b^{2}g$ ($r(t)$ is an even function and, therefore, we may take into account the whole range $-\infty < t < \infty$, comprising both signs from (3.4); in other words, the particle starts at $t = -\infty$ from $r = 0$, reaches $r_{max}$ at $t = 0$ and then falls free to the singularity when $t \rightarrow \infty$). 

As far as the angular component is concerned, we have from (3.2) 
  \begin{equation}
  \frac{d\phi}{dt} = \frac{g^{2}L}{E}
 \label{3.6}
\end{equation}
which leads to 
  \begin{equation}
  \phi(t) = \omega t,
 \label{3.7}
\end{equation}
with $\phi(0) = 0$. The constant $\omega = g^{2}L/E$ is the angular velocity of the test particle. This very simple result comes from the particular form of the metric (2.1). 

We are now in a position to write down the curve $r(\phi)$. We pick up $t(\phi)$ from (3.7) and introduce it in (3.5). Hence
  \begin{equation}
  r(\phi) = \frac{bL}{\sqrt{g^{2}L^{2} + b^{4}\phi^{2}}},
\label{3.8}
\end{equation}
where $\phi \in [0, \infty)$. The trajectory is a spiral. The motion is not periodic due to the lack of centrifugal forces keeping the particle stationary.\\
b)\textbf{ Null geodesics}\\
For the null curves we obtain from (2.1) ( $ds^{2} = 0$)
  \begin{equation}
 g^{2}r^{2} \dot{t}^{2} - \frac{1}{g^{2}r^{2}}\dot{r}^{2} - r^{2} \dot{\phi}^{2} = 0,
\label{3.9}
\end{equation}
where $\dot{t} = dt/d\lambda$, etc. and $\lambda$ is the affine parameter along the null geodesics. We have now
  \begin{equation}
  \dot{r}^{2} = b^{2},
\label{3.10}
\end{equation}
namely $\dot{r} = \pm b$ which, combined with (3.2) yields
  \begin{equation}
  \frac{1}{r(t)} = \frac{1}{r_{0}} \pm \frac{bg^{2}}{E}t
\label{3.11}
\end{equation}
with $r_{0} = r(0)$ and ($\pm$) corresponds to the ingoing and, respectively, outgoing trajectories. One observes that the condition $t \leq E/br_{0}g^{2} \equiv t_{max}$ must be obeyed, with $r(t_{max}) = \infty$. The curves $r(\phi)$ are given by 
  \begin{equation}
  \frac{1}{r(\phi)} = \frac{1}{r_{0}} \pm \frac{b}{L}\phi
\label{3.12}
\end{equation}
 where we have a $\phi_{max} = L/br_{0}$ for the outgoing geodesics and $r(\phi_{max}) = \infty$.
 
 \section{Cylindrically-symmetric case}
 Let us take now into consideration the cylindrically symmetric version of the metric (2.1)
   \begin{equation}
ds^{2} = -g^{2}r^{2} dt^{2} + \frac{dr^{2}}{g^{2}r^{2}} + dz^{2} + r^{2} d\phi^{2}
\label{4.1}
\end{equation}
where $(r, \phi)$ represents here the polar coordinates. The $z = 0$ hypersurface of (4.1) corresponds physically to $\theta = \pi/2$ hypersurface of (2.1). The geometric features here are very different compared to those of (2.1). We found that the curvature invariants are regular everywhere. Moreover, they are constant: $R^{a}_{~a} = -6g^{2},~K = 12 g^{4}$ and the Weyl tensor is vanishing (see also \cite{HC1}). We have again a horizon at $r = 0$ (that is no longer a singularity) and the same expression for the acceleration (2.5) of a static observer. In contrast, the source of curvature comes from a stress tensor with constant negative energy density and positive pressures
   \begin{equation}
  \rho = -\frac{g^{2}}{8\pi} = -p_{r} = -p_{\phi},~~~ p_{z} = \frac{3g^{2}}{8\pi}
\label{4.2}
\end{equation}
 From the expressions (4.2) we observe that the energy conditions WEC and DEC are not satisfied ($\rho < 0$) but the NEC and SEC are always valid. 
 As far as the TK energy is concerned, one obtains
 \begin{equation}
W = \frac{g^{2} r^{2}}{2} \Delta z
\label{4.3}
\end{equation}
where an integration over a cylinder of radius $r$ and length $\Delta z = z_{2} - z_{1}$ was performed. Even though $\rho$ is negative, $W$ is, nevertheless, positive due to the positive pressures. 

We would like now to estimate the energy $W$ enclosed by a cylinder with $r = 1m,~\Delta z = 1m$ and $g = 10 m/s^{2}$. One obtains $W = (g^{2}r^{2}/2G) \Delta z \approx 10^{12} J$. For the energy density one obtains $\rho = -(c^{2}/8\pi G)g^{2} \approx -5.10^{27} J/m^{3}$.

\section{Regularized spacetime}
Eqs. (2.4) shows us that the metric (2.1) is singular at $r = 0$. We look now for a regular form to render it finite everywhere. In the spirit of \cite{HC2, HC3}, we add new terms at the metric coefficients of (2.1) and obtain
  \begin{equation}
ds^{2} = -(1 + g^{2}r^{2} - e^{-\frac{1}{gr}}) dt^{2} + \frac{dr^{2}}{1 + g^{2}r^{2} - e^{-\frac{1}{gr}}} + r^{2} d\Omega^{2},
\label{5.1}
\end{equation}
where $f(r) \equiv 1 + g^{2}r^{2} - e^{-\frac{1}{gr}} > 1$ for any $r$ and $f \rightarrow 1$ when $r \rightarrow 0$. 

 The exponential term in (5.1) replaces the monopole charge $\eta^{2}$ from \cite{ND2}. It plays the role of a regulator, to make the curvatures and the stress tensor finite throughout the spacetime. The sources of (5.1) are given by
   \begin{equation}
   \begin{split}
  \bar{T}^{t}_{~t} = -\bar{\rho} = 3g^{2} - \frac{1 + gr}{gr^{3}}e^{-\frac{1}{gr}},~~~ \bar{T}^{r}_{~r} = \bar{p_{r}} = -\bar{\rho}, \\  \bar{T}^{\theta}_{~\theta} = \bar{T}^{\phi}_{~\phi} = 3g^{2} - \frac{1}{2g^{2}r^{4}}e^{-\frac{1}{gr}} \equiv \bar{p}_{t},
\label{5.2}
\end{split}
\end{equation}
where $\bar{\rho}$ is the energy density of the anisotropic fluid and $\bar{p}_{t}$ are the transverse pressure. 

It is an easy task to find that $\bar{\rho} < 0$ and $\bar{p}_{t} > 0$ for any $r > 0$. Even though the functions $\bar{\rho}(r)$ and $\bar{p}_{t}(r)$, like $f(r)$, are not analytic at $r = 0$, they however acquire finite values at the origin, i.e. $\bar{\rho} \rightarrow -3g^{2}$ and $\bar{p}_{t}, \bar{p}_{r} \rightarrow 3g^{2}$ when $r \rightarrow 0$. In addition, the same values are reached at infinity. While $\bar{\rho}$ has a maximum at $r = (\sqrt{3}-1)/2g$, $\bar{p}_{t}$ attains its minimum value $3g^{2}(1 - 128/3e^{4})$ at $r = 1/4g$. We also notice that, for $r >> 1/g$, $f(r) \approx g^{2}r^{2}$ and (5.1) becomes (2.1). Moreover,  $\bar{T}^{a}_{~b}$ acquires a $\Lambda$ - form, namely $\bar{T}^{a}_{~b} \propto \Lambda \delta^{a}_{b}$. This is consistent with (2.2) ($gr >> 1$), when the spacetimes (2.1) and (5.1) become AdS ($\Lambda = -3g^{2}$). 

As long as the scalar curvature is concerned, it is finite, with
   \begin{equation}
   R^{a}_{~a} = -12g^{2} + \frac{1 + 2gr + 2g^{2}r^{2}}{g^{2}r^{4}} e^{-\frac{1}{gr}}.
\label{5.3}
\end{equation}
 We have also $R^{a}_{~a} \rightarrow -12g^{2}$ when $r \rightarrow 0$ and at infinity. A similar behavior has the Kretschmann scalar $K$, i.e. $K = 24g^{4}$ at the origin and at infinity.
 
 Let us see now what energy conditions are satisfied by the energy-momentum tensor  $\bar{T}^{a}_{~b}$. The WEC and NEC are, of course, not obeyed because $\bar{\rho} < 0$. We have
   \begin{equation}
  \bar{\rho} + \bar{p}_{t} = \frac{-1 + 2gr + 2g^{2}r^{2}}{2g^{2}r^{4}} e^{-\frac{1}{gr}},
\label{5.4}
\end{equation}
which is positive only for $r \geq (\sqrt{3}-1)/2g$, the domain in which the NEC and SEC are satisfied. 

To take a look on some kinematical quantities, we investigate a congruence of static observers in the geometry (5.1), where the velocity vector field is given by
  \begin{equation}
   u^{b} = \left(\frac{1}{\sqrt{1 + g^{2}r^{2} - e^{-\frac{1}{gr}}}}, 0, 0, 0\right)
\label{5.5}
\end{equation}
The non-zero radial acceleration reads
  \begin{equation}
   a^{r} = g^{2}r \left(1 - \frac{1}{2g^{3}r^{3}} e^{-\frac{1}{gr}}\right) \equiv \frac{1}{2} f'(r)
\label{5.6}
\end{equation}
Simple calculations show that $a^{r} > 0$ for any $r > 0$. It vanishes when $r \rightarrow 0$ but diverges at infinity, where the exponential term becomes unity and $f(r) \approx g^{2}r^{2}$. Therefore, the gravitational field is attractive. Nevertheless, $a^{r}(r)$ is not a monotonic function. It has two extrema, a local maximum close to the origin and a minimum somewhere between $1/4g$ and $1/3g$. In addition,  $a^{r}(r)$ has an oblique asymptote  $a^{r} = g^{2}r$ at infinity and $a'(r) \rightarrow g^{2}$ when $r \rightarrow 0$.

The TK energy $W(r)$ may be computed by means of the formula (2.8) which yields
  \begin{equation}
   W = g^{2}r^{3} \left(1 - \frac{1}{2g^{3}r^{3}} e^{-\frac{1}{gr}}\right). 
\label{5.7}
\end{equation}
One notices that $W \rightarrow 0$ at the origin and becomes divergent at infinity. As in the previous non-regular case (Eq. 2.7), $W$ may be written as $W = r^{2}a^{r}$, which is positive for any $r$. Moreover, $W(r)$ is not monotonic. We have two extrema that could be found from
  \begin{equation}
   W'(r) = (r^{2}a^{r})' = 2ra^{r} + r^{2}(a^{r})' = 3g^{2}r^{2} \left(1 - \frac{1}{6g^{4}r^{4}} e^{-\frac{1}{gr}}\right). 
\label{5.8}
\end{equation}
 The above equation is transcendental and it cannot be solved analytically. Nevertheless, one notes that a necessary condition to have real roots is $(a^{r})' < 0$, that is they are located between the two extrema of the function $a^{r}(r)$ from (5.6).

 \section{Conclusions}
The curious Dadhich metric with no centrifugal radial repulsion is analysed in this paper. We looked for a different interpretation of the stresses giving rise to such a spacetime, compared to the global monopole introduced by Dadhich. Contrary to the constancy of the transverse pressures, the energy density $\rho = - p_{r}$ and they switch their signs at a definite value $r = 1/\sqrt{3}g$. They are also singular at $r = 0$ due to the divergence of the scalar curvature. It is worth to mention that the Tolman-Komar energy vanishes at the singularity because of the pressures' contribution. It is interesting to observe that the timelike and null geodesic particles have constant angular velocity $\omega = g^{2}L/E$ and the curves $r(\phi)$ are spirals.

 We modified the metric (2.1) to render it regular at the origin thanks to an exponential term added at the line-element. The energy density $\rho$ and all pressures are negative and, respectively, positive for any $r > 0$ and reach the values $\mp{3g^{2}}$ at $r = 0$.

For the cylindrically symmetric situation the stress tensor is constant, with a negative energy density and positive pressures. In addition, the horizon at $r = 0$ is no longer a singularity and the curvature invariants are constant.

\end{document}